\documentclass[
  journal=jctcce,
  manuscript=article,
]{achemso}

\usepackage{braket}
\usepackage{graphicx}
\usepackage{amsmath}
\usepackage{color}
\usepackage{natbib}

\author{Sheng Guo}
\affiliation{Department of Chemistry, Princeton University, Princeton, New Jersey 08544, USA}
\author{Mark A. Watson}
\affiliation{Department of Chemistry, Princeton University, Princeton, New Jersey 08544, USA}
\author{Weifeng Hu}
\affiliation{Department of Chemistry, Princeton University, Princeton, New Jersey 08544, USA}
\author{Qiming Sun}
\affiliation{Department of Chemistry, Princeton University, Princeton, New Jersey 08544, USA}
\author{Garnet Kin-Lic Chan}
\affiliation{Department of Chemistry, Princeton University, Princeton, New Jersey 08544, USA}
\email{gkchan@princeton.edu}

\title[DMRG-SC-NEVPT2]{$N$-electron valence state perturbation theory based on a density matrix renormalization group reference function, with applications to the chromium dimer and
poly-p-phenylene vinylene oligomer}

\begin{document}

\begin{abstract}

The strongly-contracted variant of second order $N$-electron valence state perturbation theory (NEVPT2) is an efficient perturbative method to
treat dynamic correlation without the problems of intruder states or level shifts, while the density matrix renormalization group (DMRG)
provides the capability to tackle static correlation in large active spaces.
We present a combination of the DMRG and strongly-contracted NEVPT2 (DMRG-SC-NEVPT2) that uses an efficient algorithm to compute high order
reduced density matrices from DMRG wave functions. The capabilities of DMRG-SC-NEVPT2 are demonstrated on calculations of the chromium dimer
potential energy curve at the basis set limit, and the excitation energies of poly-p-phenylene vinylene trimer (PPV(n=3)).

\end{abstract}

\section{Introduction}

In many quantum chemistry applications, such as to excited states, non-equilibrium geometries,
and transition-metal chemistry, the mean-field description of electron interactions is insufficient.
The deviation from mean-field behaviour of the electrons, i.e. electron correlation, is usually classified
into two types: static or non-dynamic correlation, and dynamic correlation.
Conventionally, static correlation is the correlation between near-degenerate valence orbitals, and can be treated by
considering the set of possible configurations of electrons in the set of active valence orbitals.
If all such configurations are included and the orbitals are simultaneously optimized, one arrives at the complete active space
self-consistent-field (CASSCF) model\cite{das1972new, werner1980quadratically, werner1985second}, the most widely used method to describe static correlation.
Dynamic correlation is the correlation between valence orbitals and the remaining empty (virtual) or doubly occupied (core) states.
This effect is thought of
as captured by low particle-rank excitations, such as used in
perturbation theories (PT)\cite{bartlett1978many}, configuration interaction singles and doubles (CISD)\cite{shavitt1977method, purvis1982full}, or coupled cluster theory (CC)\cite{coester1958bound, taylor1978unlinked, bartlett1978many, pople1978electron, chiles1981electron, bartlett1989coupled}.
The combined treatment of static and dynamic correlation is the domain of multireference correlation theories, such as
multi-reference perturbation theories (MRPT)~\cite{andersson_second-order_1990}, multi-reference configuration interaction (MRCI)~\cite{lischka1981new, szalay2011multiconfiguration, lischka2001high} or multi-reference coupled cluster theories (MRCC)~\cite{MRCC_1999, MRCC_2009, MRCC_2010}.

The density matrix renormalization group (DMRG) \cite{white_density_1992, white_density-matrix_1993, ostlund_thermodynamic_1995, rommer_class_1997, white_ab_1999, Mitrushenkov2001, chan_highly_2002, legeza_controlling_2003, Legeza2003, legeza2003optimizing, legeza2004quantum, Chan2004, moritz2005convergence, moritz2005relativistic, rissler2006measuring, Zgid2009, legeza2008applications, Reiher2010review, Chan2011, schollwock_density-matrix_2005, schollwock_density-matrix_2011} has made it possible to employ large active spaces
to describe static correlation\cite{white_ab_1999, chan_highly_2002, legeza_controlling_2003, moritz_decomposition_2007, verstraete_matrix_2008, marti_density_2008, zgid_spin_2008, kurashige_high-performance_2009, luo_optimizing_2010, chan_density_2011, marti_new_2011, sharma_spin-adapted_2012, wouters_longitudinal_2012, wouters_thouless_2013, kurashige_entangled_2013, wouters_communication:_2014} and it has become straightforward to
use the DMRG as a robust numerical solver in a variety of frontier applications, including benchmark solutions of small molecules\cite{chan_highly_2002, legeza_controlling_2003, luo_optimizing_2010, olivares-amaya_ab-initio_2015}, multi-center transition metal clusters\cite{kurashige_high-performance_2009, marti_new_2011, kurashige_entangled_2013, sharma_low-energy_2014}, as well as to molecular crystals\cite{yang_ab_2014}. There have been several
efforts to further include dynamic correlation on top of the DMRG reference.
For example, DMRG-CT~\cite{neuscamman_review_2010, Yanai2010}, DMRG-CASPT2 ~\cite{kurashige_second-order_2011}, DMRG-cu(4)-MRCI ~\cite{saitow_fully_2015} (in an internally contracted (IC) formulation) and MPS-PT~\cite{sharma_communication:_2014} and MPS-LCC ~\cite{sharma_multireference_2015} (in an uncontracted
formulation)
are all attempts in this direction. There have also been efforts to combine DMRG with density functional theory (DFT) ~\cite{reiher_density_2015}.
In the internally contracted multireference methods, high order reduced density matrices (RDM) of the DMRG active space wavefunction are required.
Such density matrices, although simple in a theoretical sense, are computationally non-trivial to obtain.

In this work, we have developed a general algorithm to efficiently compute high order reduced density matrices from DMRG wave functions.
Using these RDM calculations, we may then implement a variety of methods to treat electron correlation external to the DMRG active space.
As an example of this, we have implemented the strongly contracted variant of second order $N$-electron valence state perturbation theory (NEVPT2)
\cite{angeli_introduction_2001,angeli_n-electron_2001, angeli_n-electron_2002}, an intruder-state-free multireference perturbation theory, on top
of the DMRG reference wavefunction, to yield a method we call DMRG-SC-NEVPT2.
To demonstrate the potential of this approach, we apply the method to study the potential energy curve of the chromium dimer, and
to compute the excitation energies of the trimer of the quasi-one dimensional conjugated poly(p-phenylene vinylene) (PPV(n=3)).
The chromium dimer is a particularly demanding small molecule that has been widely studied with many techniques\cite{roos_ground_2003,celani_cipt2_2004,angeli_third-order_2006,muller_large-scale_2009,kurashige_second-order_2011,ruiperez_complete_2011,kurashige_multireference_2014,sharma_multireference_2015}, and thus forms a good test bed to demonstrate the performance of DMRG-SC-NEVPT2. Using an active space with 22 orbitals and basis sets extrapolated to the complete basis set (CBS) limit, we show that we can
compute both the spectroscopic constants and the full curve,
with an accuracy that compares quite favourably to earlier calculations.
PPV on the other hand, is of interest as a prototypical light-emitting polymer\cite{burroughes_light-emitting_1990,friend_electroluminescence_1999, barford_book}.
Light emission relies
on the correct ordering of the first optically bright $1^{1}B_{u}$ state
and  dark $2^{1}A_{g}$ state, and describing this ordering requires an appropriate
treatment of electron correlation~\cite{beljonne_theoretical_1995,lavrentiev_theoretical_1999,shukla_correlated_2002, han_time-dependent_2004, saha_investigation_2007, bursill_symmetry-adapted_2009}.
With DMRG-SC-NEVPT2, we show that we can compute the excitation
energies and energy ordering of the different states accurately, starting
from the full-valence active $\pi$ space of the molecule.

\section{Theories}

\subsection{DMRG wavefunction and optimization algorithm}

As with other wavefunction methods in quantum chemistry, DMRG is based on a wavefunction ansatz, namely the matrix product state (MPS).
An MPS is a non-linear wavefunction, built from a contraction of tensors for each orbital in the basis. If an MPS is
composed of tensors with a limited dimension (called the bond dimension, $M$) it explores only a physically motivated, but restricted, subset of the Hilbert space.
By increasing $M$, the MPS ansatz will then converge to the full configuration interaction (FCI) result. The DMRG algorithm provides a combination of
renormalization and truncation steps that efficiently find a variational and optimal MPS.
In practice, two slightly different types of MPS are used in practical
DMRG calculations: the one-site and two-site MPS.

The one-site MPS is defined as
\begin{equation}
  \ket{\Psi}= \sum_{\substack{n_1, n_2\cdot\cdot\cdot n_p\cdot\cdot\cdot n_k}} {\bf A}^{n_1}{\bf A}^{n_2}\cdot\cdot\cdot {\bf A}^{n_{p}} \cdot\cdot\cdot{\bf A}^{n_k}
\end{equation}
Here, $n_i$ is the occupacy of orbital $i$, one of $ \{\ket{}, \ket{\uparrow}, \ket{\downarrow}, \ket{\uparrow\downarrow}\}$, and $k$ is the number of orbitals.
For a given $n_i$, ${\bf A}^{n_i}$ is  an $M\times M$ matrix, except for the first and last ones; the dimensions of ${\bf A}^{n_1}$ and ${\bf A}^{n_k}$ are $1\times M$ and $M\times 1$ respectively.
This ensures that for a given occupancy string $n_1n_2\cdots n_k$  the ${\bf A}^{n_1}{\bf A}^{n_2}\cdots{\bf A}^{n_k}$ product yields a scalar,
namely, the coefficient of the determinant $\ket{n_1n_2\cdots n_k}$.

Similarly, the two-site MPS is defined as
\begin{equation}
  \ket{\Psi}= \sum_{\substack{n_1, n_2\cdot\cdot\cdot n_p\cdot\cdot\cdot n_k}} {\bf A}^{n_1}{\bf A}^{n_2}\cdot\cdot\cdot {\bf A}^{n_{p}n_{p+1}} \cdot\cdot\cdot{\bf A}^{n_k}
\end{equation}
The only difference with the one-site MPS, is that one of the tensors (${\bf A}^{n_p n_{p+1}}$) is associated with two sites at a time. The two-site MPS thus
has slightly more variational freedom than the one-site MPS, and further
exists in different non-equivalent forms, depending on which sites are chosen for the special two-site tensor.


In the DMRG sweep algorithm, $\bra{\Psi}H\ket{\Psi}/\braket{\Psi|\Psi}$ is variationally optimized, and  a single tensor ($\mathbf{A}^{n_i}$ in a one-site MPS or $\mathbf{A}^{n_i, n_{i+1}}$ in a  two-site MPS) is updated at site $i$ during the sweep. This update leads to solving an eigenvalue problem of an effective ``superblock'' Hamiltonian,
which is a $4M^2\times 4M^2$ matrix.
The multiplication between this effective Hamiltonian and a MPS is usually computed using $O(k^2)$ ``normal'' and ``complementary'' operators
built from different sets (blocks) of the active orbitals. The cost of this eigenvalue problem is $O(k^2M^3)$ for a single tensor update, and $O(k^3M^3)$
for a whole sweep optimization.

As discussed, a two-site MPS has more variational freedom than a one-site MPS at the site where the special tensor is placed. This allows
the DMRG optimization to change quantum numbers, i.e. conserved local symmetries, during the sweep, and thus helps avoid local minima. Further,
the discarded weight associated with the renormalization step in a two-site MPS sweep optimization can be used to extrapolate away the residual
error due to finite bond dimension $M$.
However, as the two-site MPS wavefunction at each step of the sweep
is slightly different (because the special tensor is moved along the sweep),
the two-site formalism does not provide a unique definition of the DMRG
expectation values.
Thus, to evaluate the reduced density matrix, we complete the DMRG
sweep optimization in the one-site representation, to obtain a single
DMRG wavefunction. This is
referred to as the ``two-site to one-site'' algorithm (a two-site MPS optimization followed by a one-site one)~\cite{olivares-amaya_ab-initio_2015}.
From here on when we refer to an MPS, we will mean the one-site MPS.

\subsection{Reduced density matrix (RDM) evaluation algorithm}

High order RDM's are common in post-active space methods that are used
to capture the ``dynamic'' correlation outside of the active space, such as multireference configuration interaction\cite{buenker_individualized_1974} and multireference perturbation theory\cite{andersson_second-order_1990, angeli_n-electron_2002}.
The expression of quantities in terms of RDM's is a feature of the internally contracted formulations,
which avoid the expansion of the external correction to the active space wavefunction explicitly in terms of many-particle basis states (such as Slater determinants
or configuration state functions). Internally contracted CASPT2\cite{andersson_second-order_1990} and NEVPT2\cite{angeli_n-electron_2002} (both strongly
and partially contracted variants) require the four-particle RDM, while
 higher levels of correlation can require even higher order RDM's.

In principle, the evaluation of RDM's in DMRG is straightforward; the evaluation of a matrix element such as $\langle \Psi|a^\dag_i a_j |\Psi\rangle$ where $\Psi$
is an MPS, is an operation of $O(k M^3)$ cost and can be carried out
within any standard DMRG implementation. However, a naive algorithm which
evaluates each RDM element separately would lead to far too high a computational cost.
Instead, the efficient evaluation of the RDM's of an MPS requires forming suitable DMRG renormalized operators which can be reused in {\it all} RDM elements $\langle a^\dagger_i a_j \rangle$.
Such efficient algorithms to build the 2-RDM were proposed some years ago by several groups~\cite{ghosh_orbital_2008,zgid_obtaining_2008}, and require only $O(k^3 M^3)$ cost, as opposed to the $O(k^5 M^3)$
cost arising in a naive element by element evaluation. The efficient algorithm to build the 3-RDM was introduced by Kurashige and Yanai.~\cite{kurashige_second-order_2011}
Here we introduce a more general algorithm that automatically organises the necessary DMRG renormalized operators for any order of RDM.

\subsubsection{General loop over orbital type patterns and number patterns}

To organize the efficient and non-redundant evaluation of the RDM elements, we label each element in terms of two classifiers: its ``type pattern'' (which describes
the pattern of creation and annihilation operators, ordered by the sites on the lattice), and its ``number pattern'' (which describes how to partition the creation and annihilation operators).
The complete set of RDM elements can be then be obtained efficiently with
proper reuse of intermediates in a DMRG sweep, by looping over all
type patterns, and all indices corresponding to given number patterns, at each step of the sweep (with a few restrictions associated with edge cases near the ends of the sweep, described in the appendix), building the DMRG renormalized operators corresponding to the given type and number patterns, and forming
the appropriate contractions with the DMRG renormalized wavefunction.

The type pattern is precisely defined as follows. Consider an RDM element $\langle {a}^\dagger_i{a}^\dagger_j\dots {a}_m{a}_n\dots \rangle$.
We reorder the indices such that the indices of the creation and annihilation operators are arranged successively in the order as they appear
in the DMRG site ordering: ${o}_{i'}o_{j'}\dots o_{m'}o_{n'}$, with $i'\le j'\le \dots \le m' \le n'$ and where $o$ is either $a$ or $a^\dagger$.
The corresponding binary pattern of 2$N$ (where $N$ is the rank of the RDM) creation and annihilation operators appearing in the above operator string
is the type pattern of the RDM matrix element.

\begin{figure}
  \includegraphics[width=8cm]{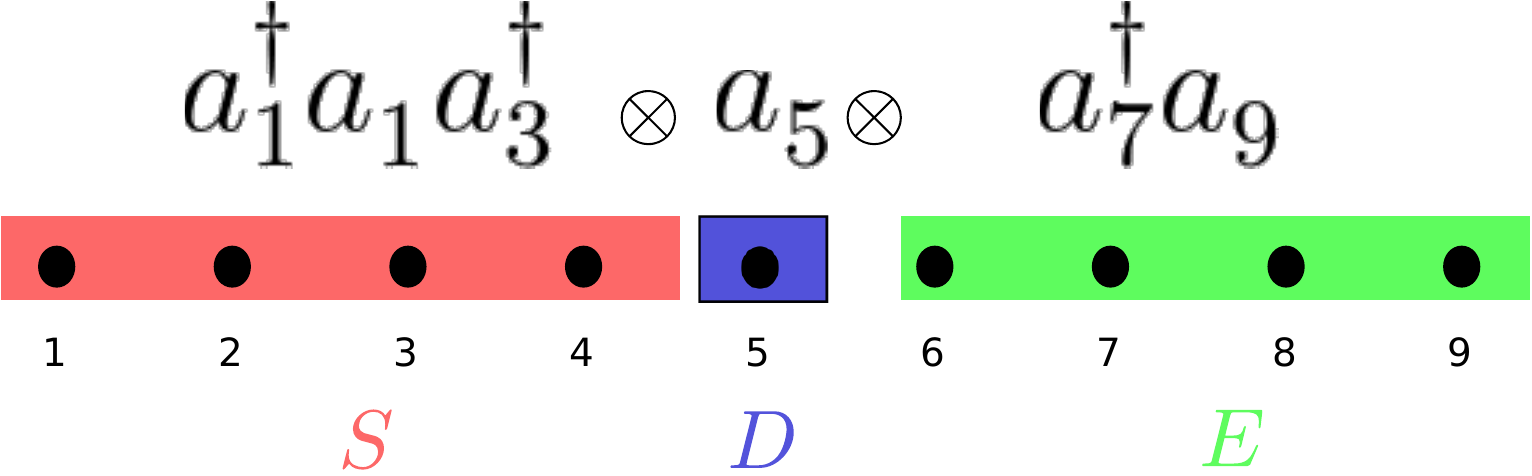}
  \caption{\label{fig:operator_split} Evaluation of a three RDM element $\gamma_{137;159}$. $a_1^\dagger a_1 a^\dagger_3$ is on $S$, $a_5$ is on $D$, and $a^\dagger_7a_9$ is on $E$. This elements belongs to the (3,1,2) ``number pattern'' and the ($a^\dagger a a^\dagger a a^\dagger a$) ``type pattern''.}
\end{figure}

The number pattern encodes how to partition the labels of the expectation value in $\langle a^\dagger_i a^\dagger_j\dots a_m a_n\dots \rangle$ onto different blocks in the DMRG sweep, so as to maximize the reuse of DMRG renormalized operators.
Recall that a DMRG sweep consists of a sequence of operations over the orbitals; at each step in the DMRG sweep, there is the set of
orbitals preceding the current orbital (the system block (${S}$)), the current orbital (the dot block (${D}$)), and the set of orbitals whose
associated site tensors have yet to be optimized (the environment block (${E}$)).
The intermediates in the construction of an RDM matrix element are DMRG renormalized operators  built and stored on the three blocks.
To determine the renormalized operators to build, we partition the $2N$ orbital labels of the operator ${O}$
among the 3 blocks. The most efficient partitioning (when generating
the full set of RDM elements) is to compute a given element from renormalized operators with
the maximum number of orbital labels appearing on the dot block, and an
equal number of  orbital labels on the system and environment blocks.
For typical RDM elements (i.e. not for elements such as $\gamma_{0000}$), at least one orbital label can be placed on $D$ and at most $N-1$ orbital labels on $E$ by choosing an appropriate position of the dot site (i.e. the step in the DMRG sweep).
The numbers \{$n_S$,$n_D$,$n_E$\} then denote the numbers of orbital labels distributed on the different blocks, and this
is the number pattern. One can generate all valid number patterns by considering all cases where the number of orbital labels on $S$ is no more than $N$, that on $D$ is no more than 4 and that on $E$ is no more than $N-1$, and the total is $2N$ except for some edge cases for the first step and last step in the sweep, as discussed in the appendix.
The number pattern and type pattern for a representative RDM element is illustrated in Fig. \ref{fig:operator_split}.

With the above process, the renormalized operators that need to be constructed  on each block $S$, $D$, and $E$
in the DMRG sweep, are enumerated
by the corresponding set of type patterns and number patterns of the RDM that is being built. In total this leads to $O(k^N)$ renormalized
operators that need to be built, most of which are constructed on the system block $S$ at different steps in the sweep. The cost to build these operators
is $O(k^NM^3)$ at each step of the sweep and $O(k^{N+1}M^3)$ in total.

\subsubsection{Computing the expectation values}

After building the renormalized operators needed on the three blocks, $[O]^S$, $[O]^D$, $[O]^E$ at site $i$ of the sweep, they are
traced with the renormalized DMRG wavefunction, i.e. the site tensor $\mathbf{A}^{n_i}$.
This corresponds to the computation
\begin{align}
\gamma = \sum_{snes'n'e'} A^{n_i}_{se} p [O]^S_{ss'} [O]^D_{nn'} [O]^E_{ee'} A^{n'_i}_{s'e'}
\end{align}
where the above tensor contraction is performed in stages (e.g. contracting $s$, $n$, $e$ indices separately) and intermediates such as $\sum_{snes'n'e'}A^{n_i}_{se}[O]^S_{ss'}A^{n'_i}_{s'e'}$ are reused to minimize the computational cost,
and $p$ is a parity operation that inserts appropriate minus signs for fermions~\cite{chan_highly_2002}.
With appropriate intermediates, the total cost of this step for an $N$-RDM computation is $O(k^{N+1}M^3+k^{2N}M^2)$.

\subsubsection{Spin recoupling}

Through the above steps, an expectation value such as $\braket{{a}^\dagger_0{a}^\dagger_1{a}_2{a}_3}$ is obtained.
For a non-spin-adapted DMRG algorithm, where $a^\dag, a$ are simple creation and annihilation operators,
we can then obtain $\gamma_{0132}$ by permuting $ {a}_2$ and ${a}_3$ with appropriate signs. For spin-adapted DMRG implementations, the $a^\dag, a$
are spin tensor creation and annihilation operators, and
$\braket{{a}^\dagger_0{a}^\dagger_1{a}_2{a}_3}$ is instead an element such as
$\braket{\{[({a}^\dagger_0{a}^\dagger_1)^{S_1}_{S}({a}_2)_{D}]^{S_2}({a}_3)_{E}\}^{S_3}}$ with different spins $S_1$, $S_2$, $S_3$ and spin couplings.
The spin tensor creation and annihilation operators can be expanded as a linear combination of spin orbital creation
and annihilation operators, leading to a set of linear equations to obtain the spin orbital RDM elements from a single spin reduced matrix element. Using singlet embedding \cite{tatsuaki_interaction-round--face_2000, sharma_spin-adapted_2012}, expectation values for the spin tensor combinations with non-zero total spin are zero, significantly decreasing the number of
expectation values to compute. The coefficients appearing in the linear equations are the same for all operators with the same pattern and can thus be generated
automatically and reused. The cost of this process is independent of the DMRG bond dimension $M$, and is thus negligible compared to the other parts of the RDM calculation.

\subsection{$N$-electron valence state perturbation theory}

We now briefly review the strongly contracted NEVPT2\cite{angeli_n-electron_2001, angeli_n-electron_2002}, a second-order
multireference perturbation theory. The target zeroth-order wave function is defined as
\begin{equation}
P_{CAS}HP_{CAS} \ket{\Psi^{(0)}} = E^{(0)} \ket{\Psi^{(0)}}
\end{equation}
where $P_{CAS}$ is the projector onto the CAS space.
The zeroth order Hamiltonian is chosen in the form given by Dyall\cite{dyall_choice_1995}:
\begin{equation}
  H^D = H_c + H_v + C
\end{equation}
where $H_c$ is a one-electron (diagonal) operator in the non-active (core and external) subspace:
\begin{equation}
  H_c = \sum_{i,\sigma}^{core} \epsilon_i  a^\dagger_{i,\sigma} a_{i,\sigma} + \sum_{r,\sigma}^{virt} \epsilon_r  a^\dagger_{r,\sigma} a_{r,\sigma}
\end{equation}
Indices $(i, j,\dots)$, $(a, b,\dots)$ and $(r,s,\dots)$ refer to canonical core orbitals, active orbitals and canonical virtual orbitals, respectively, and $\epsilon_i $ and $\epsilon_r$ are orbital energies.
$H_v$ is a two-electron operator confined to the active space:
\begin{equation}
  H_v = \sum_{ab,\sigma}^{act} h^{eff}_{ab}  a^\dagger_{a,\sigma}  a_{b,\sigma} + \sum_{abcd,\sigma,\eta}^{act} \braket{ab|cd}  a^\dagger_{a,\sigma} a^\dagger_{b,\eta} a_{d,\eta} a_{c,\sigma}
\end{equation}
where $h_{ab}^{eff} = h_{ab} + \sum_{i}^{core} (2\braket{ai|bi}-\braket{ai|ib})$;
and $C$ is a constant defined as
\begin{align}
  C=2\sum_i^{core} h_{ii} + \sum_{ij}^{core}(2\braket{ij|ij} -\braket{ij|ji}) - 2\sum_i^{core}\epsilon_i
\end{align}
to ensure that $H^D$ is equivalent to the full Hamiltonian within the CAS space.

The zeroth order wave functions external to the CAS space are referred to as the ``perturber functions''. A perturber function is written  as $\ket{\Psi^{(k)}_{l,\mu}}$.
It belongs to a CAS space denoted $P_{S_l^{(k)}}$, where $k$ is the number of electrons promoted to ($k>0$) or removed from ($k<0$) the active space, $l$ denotes the occupancy of nonactive orbitals and $\mu$ enumerates the various perturber functions in $S_l^{(k)}$.
In strongly contracted NEVPT2, only one perturber function for each $S_l^{(k)}$ subspace is used. This leads
  to less variational freedom in the Hylleraas functional than partially contracted and uncontracted NEVPT2, and due to the smaller variational freedom, strongly contracted NEVPT2 will yield
  a higher value for the perturbation theory energy~\cite{angeli_n-electron_2002}. In general, however, the contribution
  of contraction to relative energies is often small on the chemical energy scale, and for recent results exploring
the effect of contraction, we refer to Ref.~\cite{alex_t_nevpt2}. The strongly contracted perturber function is defined as $\ket{\Psi^{(k)}_l} = P_{S^{(k)}_l} H \ket{\Psi^{(0)}}$, where $P_{S_l^{(k)}}$ is the projector onto the space. The energy of
the perturber function is
\begin{equation}
  E^{(k)}_l = \frac{\bra{\Psi^{(k)}_l}H^D\ket{\Psi^{(k)}_l}}{\braket{\Psi^{(k)}_l|\Psi^{(k)}_l}}
\end{equation}
The zeroth order Hamiltonian then becomes
\begin{equation}
  H_0 = \sum_{kl} \ket{\Psi^{(k)'}_l}E^{(k)}_l \bra{\Psi^{(k)'}_l} +\ket{\Psi^{(0)}}E^{(0)} \bra{\Psi^{(0)}}
\end{equation}
where $\ket{\Psi^{(k)'}_l}$ corresponds to the normalized $\ket{\Psi^{(k)}_l}$.

The bottleneck in SC-NEVPT2 is the evaluation of the energies of the perturber functions,
where up to the 4-RDM appears. This 4-RDM is subsequently contracted with
two electron integrals in the active space to form auxiliary matrices.\cite{ angeli_n-electron_2002}
Computing the 4-RDM requires $O(k^8M^2+k^5M^3)$ cost, much higher than that of a standard DMRG energy optimization.
To minimize the cost, it is generally necessary to evaluate the 4-RDM using a lower bond dimension than is used in the DMRG energy optimization of $\ket{\Psi^{(0)}}$.
To obtain an MPS approximation of lower bond dimension for the reference,
we  ``compress'' the converged reference MPS of a larger bond dimension.
 This is done in practice by carrying out a ``reverse schedule'' set of sweeps
in a DMRG code, where $M$ is decreased.~\cite{olivares-amaya_ab-initio_2015} All of our DMRG-SC-NEVPT2 calculations use reference MPS that have been
compressed in this way to compute the RDM's. The compressed bond dimension is
denoted $M'$.

Further, to avoid storage problems in our implementation of DMRG-SC-NEVPT2,
the 4-RDM is computed on the fly. In principle, it is possible to
completely avoid generating the 4-RDM by directly computing the auxiliary matrices (i.e. with the integrals already contracted).
This is the approach taken, for example, in Kurashige and Yanai's original implementation of DMRG-CASPT2
~\cite{kurashige_second-order_2011}, where the auxiliary matrices can be computed with a cost scaling similar to that of the 3-RDM,
using special kinds of DMRG renormalized operators as intermediates. However, the Dyall Hamiltonian in NEVPT2
involves all the active-space two-electron integrals, and this is more complicated to handle than the diagonal Fock energies that appear in the CASPT2 zeroth
Hamiltonian (in its canonical representation). Thus, although the scaling of such an algorithm may be lower, it is complicated to implement
and the computational prefactor is large due to the many different intermediates that need to be considered. We have implemented
such an algorithm only in our reference NEVPT2 code that uses CASCI type reference wavefunctions, and not in our DMRG-SC-NEVPT2 code.

\section{Applications}
\subsection{Chromium dimer}

The chromium dimer has been a challenging molecule for quantum chemistry for many years. This
is because a relatively large active (12$e$, 12$o$) space  is needed to
describe the minimal spin-recoupling
of the electrons involved in the nominal sextuple bond, while, at the same time, a large amount of dynamic correlation is required
even to yield a bound potential energy curve. \cite{andersson_cr2_1994,roos_multiconfigurational_1995,roos_multiconfigurational_1996,roos_ground_2003,angeli_third-order_2006,muller_large-scale_2009,kurashige_second-order_2011,sharma_multireference_2015}

The minimal (12$e$,12$o$) CAS, derived from the 3d and 4s atomic orbitals, has been  widely employed in earlier chromium dimer calculations.~\cite{andersson_cr2_1994, roos_multiconfigurational_1995, roos_multiconfigurational_1996, angeli_third-order_2006, muller_large-scale_2009, sharma_multireference_2015} Although these earlier calculations
could reproduce the general shape of the potential energy curve, they
also have a number of shortcomings.
For example, both CASPT2\cite{andersson_cr2_1994,roos_multiconfigurational_1995,roos_multiconfigurational_1996} and NEVPT2\cite{angeli_n-electron_2001} based on the CASSCF(12$e$,12$o$) reference function  significantly overestimate the
dissociation energy of the molecule, especially when large basis sets (e.g. those including g-, h-, or i- type functions) are used.\cite{celani_cipt2_2004,angeli_third-order_2006} Further, with NEVPT3, \cite{angeli_third-order_2006} starting
from the CAS(12$e$,12$o$) wavefunction  results in a large correction to the NEVPT2 result
and an unphysical potential energy curve. These various results suggest
that (at least when used with low orders of perturbation
theory and internally contracted formulations which have a limited ability to relax the reference) the minimal (12$e$,12$o$) CAS is too small for a quantitative description.

Further, when using CASPT2, the results are also  sensitive to the choice of the zeroth-order Hamiltonian and
the level shift employed.\cite{celani_cipt2_2004,ruiperez_complete_2011} This is
especially true for the (12$e$,12$o$) active space,
but sensitivity can still be seen even in  DMRG-CASPT2 calculations with a large (12$e$,28$o$) CAS derived from the 3$d$, 4$s$, 4$p$, 4$d$ atomic orbitals.~\cite{kurashige_second-order_2011}
While the large active space DMRG-CASPT2 curves are generally close to the experimentally derived
 curve, the different choices of zeroth-order level shifts
still yielded a variation in $D_e$
of about 0.2~eV.~\cite{kurashige_second-order_2011} Since NEVPT2 does not require level shifts to avoid intruders, the above provides further incentive
to recompute the chromium dimer potential energy curve using DMRG-SC-NEVPT2,
and with a large active space.

\begin{figure}
  \includegraphics[width=1.0\columnwidth]{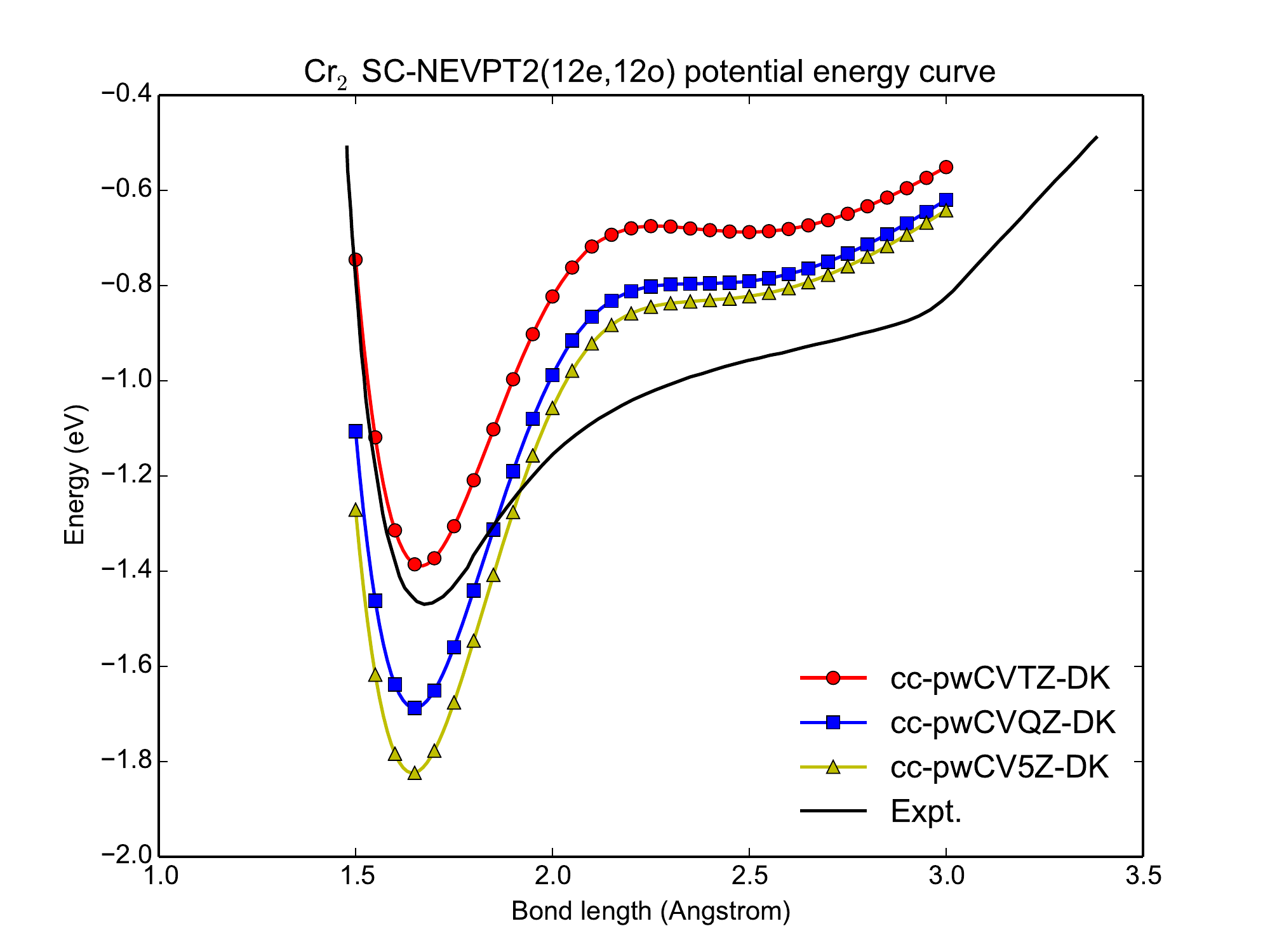}
  \caption{SC-NEVPT2(12$e$,12$o$) potential energy curve with a suite of cc-pwCVXZ-DK (X=T, Q, 5) basis sets.
The experimental curve is taken from Ref.~\cite{casey_negative_1993}.
  }
  \label{fig:12o_nevpt2}
\end{figure}

For our DMRG-SC-NEVPT2 calculations, we used a suite of cc-pwCVXZ-DK (X = T, Q, 5) basis sets\cite{balabanov_2005_systematically}. No basis set superposition error (BSSE) corrections were applied
due to the large basis sets used (which included h- and i-type functions). The X2C Hamiltonian (spin-free one-electron variant)~\cite{liu_ideas_2010, saue_relativesitic_2011, peng_exact_2012} was used to describe scalar relativistic effects,
  and to capture the relativistic core contraction,
  uncontracted basis sets (the DK variants
  of the correlation consistent basis sets) were used.
The DMRG reference and corresponding RDM and DMRG-SC-NEVPT2 calculations were performed using the \textsc{Block} program\cite{sharma_spin-adapted_2012}.
All other calculations were performed in \textsc{Pyscf} \cite{sun_pyscf}. In the figures, the energy of the isolated atoms is set to zero.

We first show the SC-NEVPT2(12$e$,12$o$) potential energy curve of Cr$_2$ in Fig.~\ref{fig:12o_nevpt2}
to demonstrate the basis set convergence with the cc-pwCVXZ-DK bases. The (12$e$,12$o$) active space was obtained from a CASSCF calculation. We find that
using larger basis sets together with this small active space indeed yields far too deep a curve, in agreement with earlier studies  using atomic natural orbital (ANO) basis sets.~\cite{angeli_third-order_2006}

We then extended the (12$e$,12$o$) active space by adding another set of $\sigma$, $\pi$, $\pi'$, $\delta$, $\delta'$ orbitals and their corresponding anti-bonding orbitals
to obtain a (12$e$,22$o$) active space. We selected the orbitals from the converged (12$e$,12$o$) CASSCF calculation based on symmetry and their orbital energies,
and the orbitals were then further optimized using DMRG-CASSCF ($M$=1000, no frozen core). Although the additional active orbitals contained both $4p$ and $4d$ character at first,
the fully optimized additional active orbitals were mainly of $4d$ character.  This is not surprising, as a $4d$ double shell is well known to greatly improve transition metal CAS results~\cite{andersson_excitation_1992}.
From the DMRG-CASSCF optimized orbitals, we then carried out a final larger DMRG calculation with $M$ = 4000 to obtain an accurate reference wavefunction.
The resulting DMRG reference energy in the active space is accurate to better than 0.1 m$E_h$.
To compute the RDM's for the DMRG-SC-NEVPT2 calculation, the large bond dimension reference MPS was compressed down to an MPS with $M'$ = 800.
Finally, the (TZ/QZ/5Z) energies were extrapolated to the CBS limit using
an exponential formula for the DMRG-CASSCF energy, and an $l^{-3}$ formula for the DMRG-SC-NEVPT2 correction, where $l$ is the maximum angular momentum
of the basis.

\begin{figure}
  \centering
  \includegraphics[width=1.0\columnwidth]{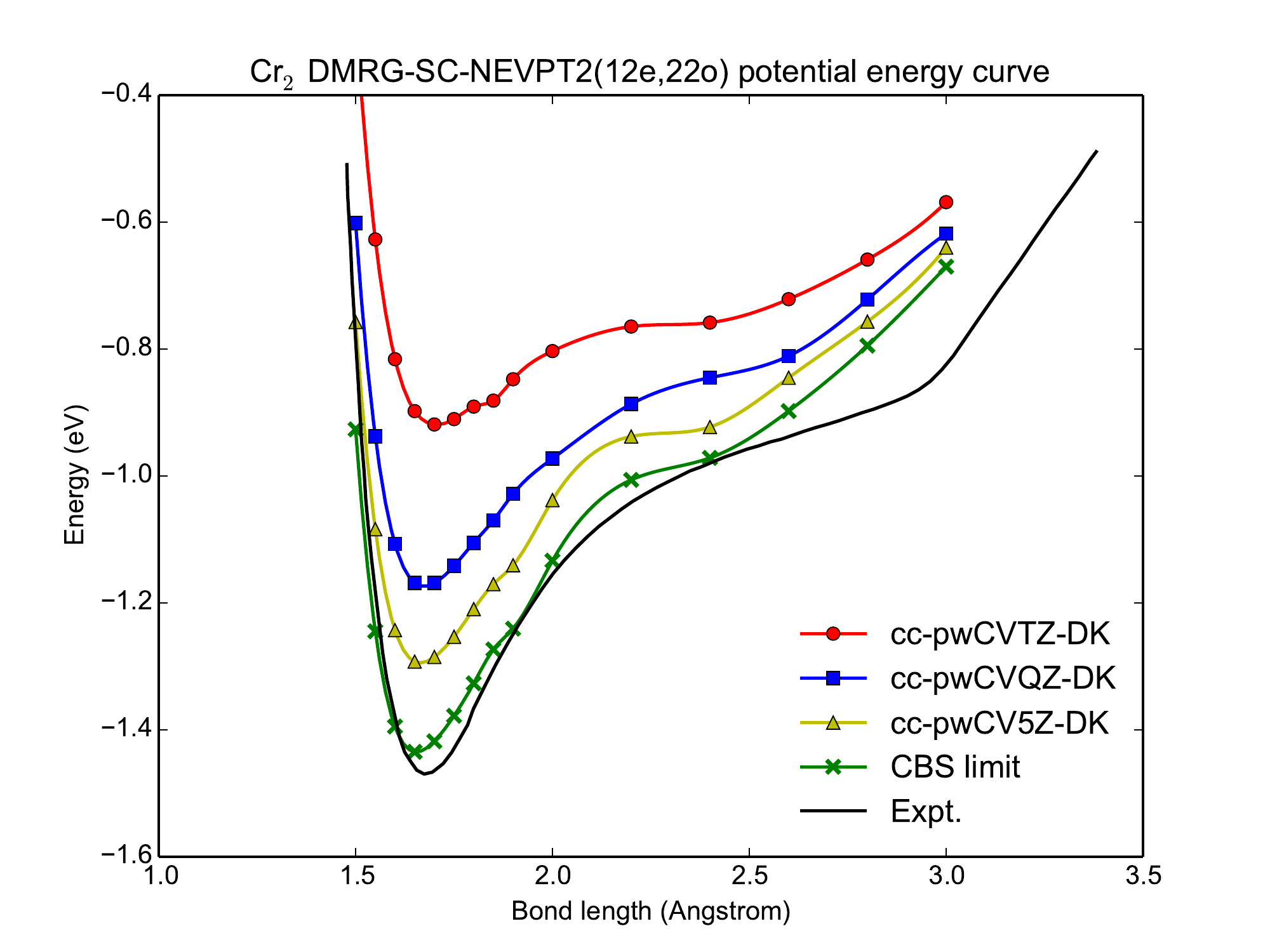}
  \caption{Extended active space DMRG-SC-NEVPT2(12$e$,22$o$)($M'$=800) potential energy curves with a suite of cc-pwCVXZ-DK (X=T, Q, 5) basis sets,
    and the extrapolated CBS limit. The extended active space includes a double $d$ shell.
The experimental curve is taken from Ref.~\cite{casey_negative_1993}.}
  \label{fig:dmrg-nevpt2}
\end{figure}
\begin{figure*}
  \includegraphics[width=1.0\columnwidth]{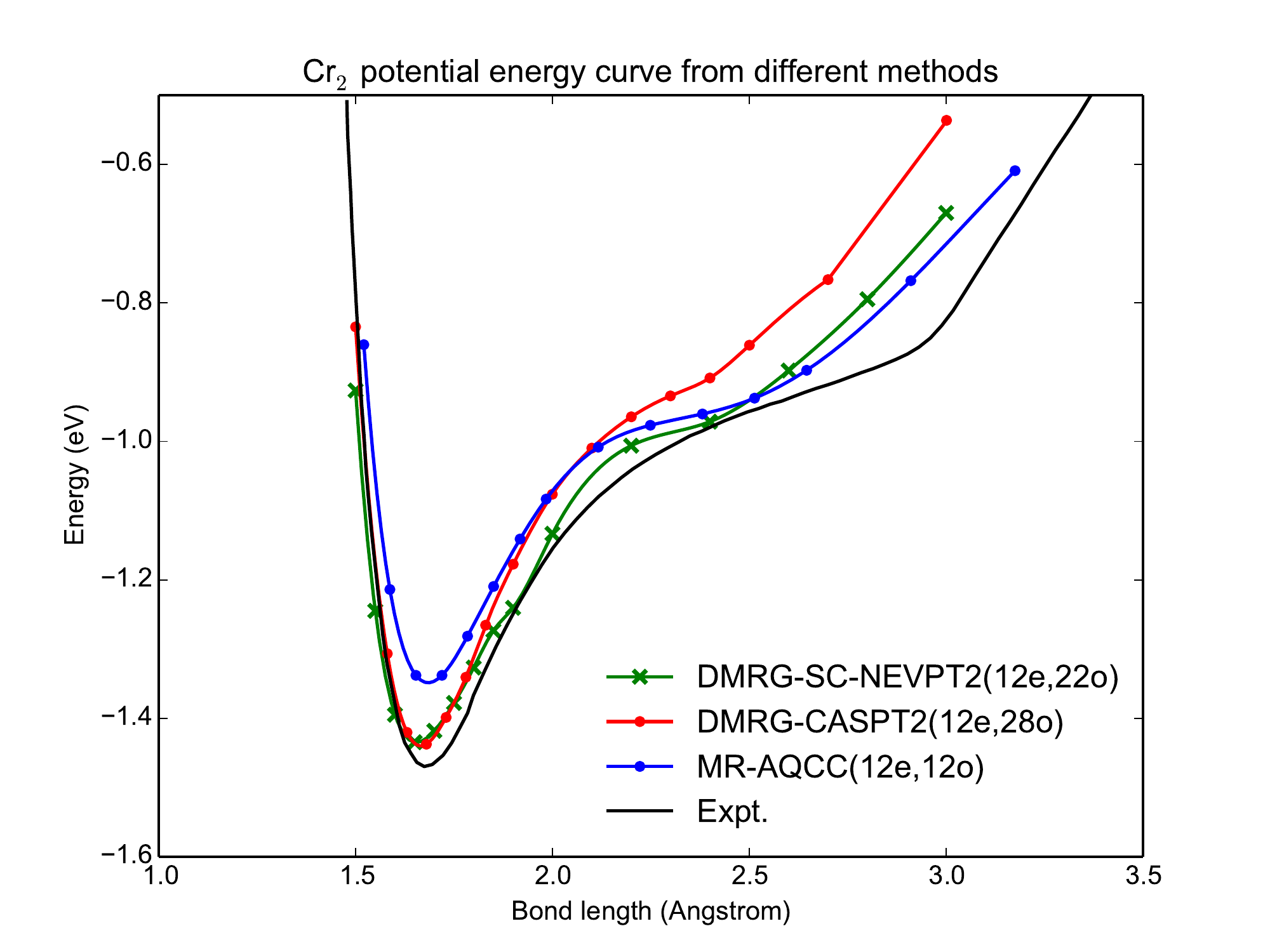}
  \caption{Cr$_2$ potential energy curves from different methods.  The DMRG-CASPT2(12e,28o) curve is from Ref.~\cite{kurashige_second-order_2011} and the MR-AQCC(12$e$,12$o$) curve is from Ref.~\cite{muller_large-scale_2009}. The experimental curve is taken from Ref.~\cite{casey_negative_1993}.
  }
  \label{fig:compare}
\end{figure*}

Figure \ref{fig:dmrg-nevpt2} shows the potential energy curves obtained with the cc-pwCVXZ-DK (X=T, Q, 5) basis sets and the X2C Hamiltonian, as well as the extrapolated
CBS curve. A comparison of the CBS curve to curves obtained with other
correlation methods is shown in Fig.~\ref{fig:compare}.
We see that larger basis sets using this active space yield systematically improved curves, and that
the CBS potential energy curve now agrees very well with the experimental curve, over a wide range of distances.
There is a small non-smoothness in the curve near 1.85\AA, of roughly 0.006 $eV$ (0.2 m$E_h$). This small
  non-smoothness reflects the limited bond-dimension of $M'=800$ used in the NEVPT2 part of the calculation, which means
  the PT2 part of the energy is converged to about 0.2 m$E_h$.
Otherwise, the only significant deviation from experiment occurs near the bend at about 2.8\AA, where the reliability of the RKR potential has previously been questioned.\cite{roos_ground_2003}
The spectroscopic constants are shown in Table \ref{tab:spectro}.
Our obtained CBS $D_e$ of 1.43eV is very close to the inferred experimental binding energy of 1.47 eV from
the negative ion photoelectron spectroscopy measurements
of Casey and Leopold~\cite{casey_negative_1993}, while the bond-length and vibrational frequency also agree well. It is remarkable to see that even
with the largest cc-pwCV5Z-DK basis, the binding energy is still 0.14 eV away from the CBS limit, demonstrating the very large basis set
effects in this system. In Sokolov and Chan's recent results using time-dependent (i.e. uncontracted) NEVPT2 in a (12$e$,12$o$) active space\cite{alex_t_nevpt2}, the effect of the strong contraction approximation amounts to $\approx$ 0.15 eV in the binding in the equilibrium region of the Cr$_2$ potential energy
  curve and decreases monotonically towards the dissociation limit. Assuming a smaller effect of the perturbation and of its contraction error in the larger active space, the small underbinding observed here might thus be improved within the uncontracted formulation.

Compared to earlier ``accurate'' calculations, the computed DMRG-SC-NEVPT2 curve
here compares  quite favourably. For example, near the equilibrium
distance, the DMRG-SC-NEVPT2(12$e$,22$o$) curve is almost identical to the DMRG-CASPT2(12$e$,28$o$) curve\cite{kurashige_second-order_2011}. However, in the stretched $4s$ bonding (``shoulder'') region, the DMRG-SC-NEVPT2(12$e$,22$o$) curve is closer to the experimental one. This seems consistent with the increased correlation strength in the $4s$ bonding region, which could be harder to describe with the simpler zeroth order Hamiltonian employed in CASPT2. Across the full range of distances, the DMRG-SC-NEVPT2 curve also appears considerably more accurate than the MR-AQCC(12$e$,12$o$) curve, likely due to the larger active space employed.

\begin{figure}
  \centering
  \includegraphics[width=1.0\columnwidth]{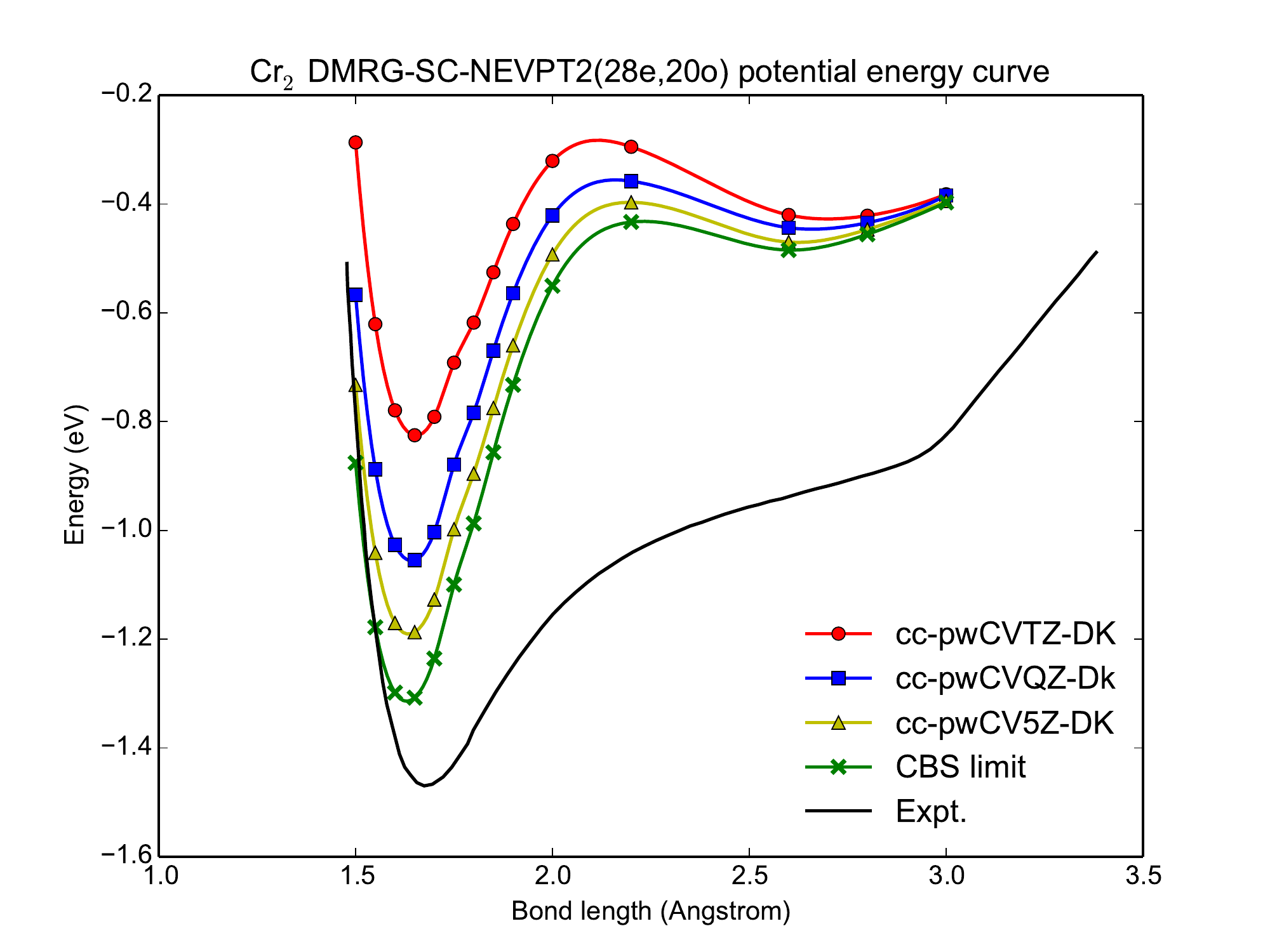}
  \caption{Extended semi-core DMRG-SC-NEVPT2(28e,20o)($M'$=800) potential energy curve with a suite of cc-pwCVXZ-DK (X=T,Q,5) basis sets, and the extrapolated CBS limit. The semi-core space includes the 3$s$ and 3$p$ shells.
The experimental curve is taken from Ref.~\cite{casey_negative_1993}. }
  \label{fig:semicore}
\end{figure}

The (12$e$,28$o$) CAS used in the earlier DMRG-CASPT2 calculations
contained $4p$ orbitals. However, the accuracy of the DMRG-SC-NEVPT2(12$e$,22$o$) curve as compared to the DMRG-CASPT2(12$e$,28$o$), together with the fact that only $4d$ orbitals remain in the active space after orbital optimization, suggests that the $4d$ orbitals are more important than the $4p$ orbitals in the correlations of this system. It has also been suggested in some studies
that semi-core correlations (involving the $3s$ and $3p$ orbitals) should also have  a large effect. In Fig.~\ref{fig:semicore} we show the corresponding curves
computed using a (28$e$,20$o$) active space containing the semi-core $3s, 3p$ orbitals and $3d$, $4s$ valence shell (but no $4d$ shell). We see that the curves are in fact much worse than the curves obtained with the (12$e$,22$o$) active space above. 
This indicates that the semi-core correlations in this system are described well at the second-order perturbation theory level and that double shell effects should be dealt
  with at a higher level of theory.

 \begin{table}
\caption{Spectroscopic constants for the ground state of Cr$_2$ using different methods.  \label{tab:spectro}}
  \begin{tabular}{cccc}
  \hline
  & $D_e(eV)$ & $R_0($\AA$)$ & $\omega_e$(cm$^{-1}$ \\
  \hline
  DMRG-SC-NEVPT2(12e,22o) $^a$&  & & \\
  $M'=$800 & 1.435 & 1.655 & 469 \\
  $M'=$1200 & 1.432 & 1.656 & 470 \\
  DMRG-CASPT2(12$e$,28$o$)$^b$ & 1.610 & 1.681 & 480 \\
  MR-AQCC(12$e$,12$o$)$^c$ & 1.355 & 1.685 & 459 \\
  FP-AFQMC$^d$ & 1.63(5) & 1.65(2) & 552(93) \\
  experiment & 1.47(5)$^e$ & 1.679$^f$ & 480.6(5)$^e$ \\
  & 1.47 $\pm$0.1$^g$& &\\
  \hline
  \end{tabular}
  \\
  $a$: From fitting the energies at five points, R=1.60,1.65, 1.70, 1.75, and 1.80\AA. The DMRG-CASSCF energies were evaluated with $M=$4000.
  $b$: DMRG-CASPT2(12$e$,28$o$)/{\bf $g_1$}($M=\infty$) with the cc-pwCV5Z-DK basis set in Ref.~\cite{kurashige_second-order_2011}
  $c$: TZP/QZP CBS limit in Ref.~\cite{muller_large-scale_2009}
  $d$: Ref.~\cite{purwanto_auxiliary_field_2015}
  $e$: Ref.~\cite{casey_negative_1993}
  $f$: Ref.~\cite{bondybey_electronic_1983}
  $g$: Ref.~\cite{su_bond_1993}
\end{table}

\subsection{Poly(p-phenylene vinylene)}

Poly(p-phenylene vinylene) is a well-known light emitting conjugated polymer. It has several low-lying states that participate in its photophysics
and its oligomers have been extensively investigated with many theoretical methods, for example various semi-empirical methods~\cite{beljonne_theoretical_1995}, DMRG based on the Pariser-Parr-Pople (PPP) model Hamiltonian~\cite{lavrentiev_theoretical_1999, shukla_correlated_2002,bursill_symmetry-adapted_2009}, time-dependent density functional theory~\cite{han_time-dependent_2004} and symmetry-adapted cluster configuration interaction (SAC-CI)~\cite{saha_investigation_2007}.
Here we examine the excited states and their energies using DMRG-SC-NEVPT2.
We limit our calculations to PPV(n=3) (three phenylene vinylene units) to demonstrate the capabilities of DMRG-SC-NEVPT2 for low-lying excited states in conjugated molecules.

The cc-pVDZ basis set~\cite{dunning1989gaussian} and a (22$e$, 22$o$) active space containing the 22 conjugated $\pi, \pi^*$ orbitals were used. The ground state equilibrium geometry of PPV(n=3) was obtained via DMRG-CI geometry optimization~\cite{hu_excited-state_2015}, which is implemented within an interface between \textsc{Block} and \textsc{ORCA}\cite{neese_orca_2012}, starting from the DFT/B3LYP optimized geometry. The DMRG-CI geometry optimization was first run at $M$ = 1000, and near convergence with $M$ = 2000. Canonical Hartree-Fock orbitals were used
in the DMRG calculations in the geometry optimization.

Compared to canonical orbitals, localized orbitals usually give better convergence in DMRG calculations.~\cite{olivares-amaya_ab-initio_2015} Therefore, further state-averaged DMRG-CI calculations were carried out for the 6 lowest singlet states, using split-localized orbitals (where the occupied and unoccupied $\pi$, $\pi*$ orbitals are separately localized with the Pipek-Mezey method~\cite{pipek_fast_1989}). The energies were converged to $0.2$ m$E_h$ with a bond dimension of $M$ = 3200. The excited states were identified according to their excitation signature from the ground state through the transition 1-RDM and 2-RDM
in the canonical Hartree-Fock basis. The symmetry of the states was also determined by the excitation signature: as the ground-state
 is an $A_g$ state, excitations between two orbitals of different symmetries lead to a $B_u$ excited state, and excitations between two orbitals of the same symmetry lead to an $A_g$ state.

Due to the lack of $\pi-\sigma$ interaction and dynamical correlation, the energy ordering of excitations in DMRG-CI is qualitatively different
from experiment, and the state with the HOMO$\rightarrow$LUMO excitation signature was not the lowest one in energy.
To obtain the correct state ordering, dynamic correlation was further included using DMRG-SC-NEVPT2. The DMRG-CI wavefunctions were compressed
to a smaller bond dimension of $M'$=500 and 750. The difference in excitation energy from these two different $M'$ bond dimensions
in the DMRG-SC-NEVPT2 calculation was less than $0.6$ m$eV$ (Table~\ref{table:local}), and thus the error from incomplete $M'$ can
effectively be ignored.
The excitation energy of the first bright state as computed by DMRG-SC-NEVPT2 ($3.86 eV$) now agrees well with
the experimental number from fluorescence in vacuo by extrapolating the linear fits to $n_{solv}=1$ ($3.69 eV$)~\cite{gierschner_fluorescence_2002} (Table~\ref{table:compare}) and the first bright ($B_u$ state) is found to lie below the $2A_g$ state, as required for efficient light emission. We also find a second $B_u$ state slightly
below the $2A_g$ state in energy. The electronic signatures of the states are shown in Table~\ref{table:local}.

\begin{table}
\caption{Vertical excitation energies (eV) of PPV(n=3) in vacuum from different methods.}
\label{table:compare}
\begin{tabular}{ccccc}
  \hline
  \hline
State  & DMRG-SC-NEVPT2 & DMRG/PPP & SAC-CI & expt\\
\hline
$1^1B_u$ & 3.86   &3.52$^a$, 3.91$^b$, 3.751$^c$, & 3.57$^d$ & 3.69$^e$ \\
$2^1A_g$ & 4.66   &4.06 $^a$, 4.618$^c$& & \\
\hline
\hline
\end{tabular}
\\
$a$: Ref.~\cite{lavrentiev_theoretical_1999}
$b$: Ref.~\cite{shukla_correlated_2002}
$c$: Ref.~\cite{bursill_symmetry-adapted_2009}
$d$: Ref.~\cite{saha_investigation_2007}
$e$: Ref.~\cite{gierschner_fluorescence_2002}
\end{table}

\begin{table*}
  \centering
  \caption{Excitation energy (eV) calculated from DMRG-SC-NEVPT2 ($M'$=500, 750) with localized orbitals.
    Excitations energies of DMRG-CI ($M$=3200) also shown for comparison. The transition signatures are calculated from the
    DMRG-CI wave function with $M$ = 3200. The excitation coefficient of the transition $i\rightarrow j$ ($i,j \rightarrow k,l$) is given by $\frac{1}{\sqrt{2}}\bra{\Psi}a^\dagger_j a_i\ket{1^1A_g}$ ( $\frac{1}{2}\bra{\Psi}a^\dagger_l a^\dagger_k a_j a_i\ket{1^1A_g}$).
The transition labels $n\rightarrow m'$ are as follows: 1, 2, 3 \ldots denote HOMO, HOMO-1, HOMO-2 \ldots canonical orbitals, while $1'$, $2'$, $3'$ \ldots denote LUMO, LUMO+1,LUMO+2 canonical orbitals. }
  \label{table:local}
  \begin{tabular}{ccccc}
    \hline
\hline
symmetry & excitation signature &  $M'$=500 & $M'$=750 & DMRG-CI($M$=3200) \\

\hline
  $1^1B_u$ & 0.83 ($1\rightarrow1')$ & 3.86   &   3.86   & 5.13 \\
  $2^1B_u$ & 0.51 ($5\rightarrow 1'$) + 0.47 (1$\rightarrow 3'$) & 4.50   &   4.50   & 4.50    \\
  $2^1A_g$ & 0.35 ($2\rightarrow 1'$) + 0.30 (1$\rightarrow 2'$) & 4.66   &   4.66   & 4.85    \\
  &   + 0.47 ($1,1\rightarrow1',1'$)  & & &\\
  $3^1B_u$ & many singly excited terms & 4.83   &   4.83   & 4.75    \\
  $4^1B_u$ & many singly excited terms & 4.84   &   4.84   & 4.76    \\
\hline
\hline
\end{tabular}
\end{table*}

\section{Conclusions}

In this work we described an efficient and general algorithm to compute high order RDM's from a DMRG wave function.
These RDM's may be used as the starting point for many kinds of internally contracted dynamic correlation methods.
Using up to the 4th order RDM, we combined DMRG and strongly contracted NEVPT2 to obtain
an intruder state free second order multireference perturbation theory that can be used with large active spaces.

To demonstrate the capability of the DMRG-SC-NEVPT2 method, we calculated the potential energy curve of the chromium dimer and the excited
states of PPV(n=3). For the chromium dimer, the best extended active space included the so called ``double d'' shell.
Our obtained curve compares quite favourably with earlier calculations and tracks the experimental curve very closely, except
at very long bond-lengths where the experimental inversion may be suspect. We find that the semi-core orbitals
are much less important for an accurate description of this curve.
For PPV(n=3), using an active space containing the full set of $\pi, \pi^*$ orbitals, we found that the dynamic correlation included through NEVPT2 allowed us to obtain
a correct ordering of the low-lying excited states, and the lowest excitation energy is in good agreement with
the experimental data. In summary, these examples paint an optimistic picture for the potential of applying the
DMRG-SC-NEVPT2 method developed here to a wide range of challenging problems.

\section*{Apppendix A: Edge cases for number patterns in RDM}
All number patterns in the $N$-RDM, \{$n_S$, $n_D$, $n_E$\}, need to be computed if
\begin{align}
0\le n_S \le N \\
0\le n_E \le N-1 \\
1\le n_D \le 4 \\
n_S +n_D +n_E =2N
\end{align}

However, there are some RDM elements, whose indices cannot be arranged as described above, because too many of the indices lie on the
first or last site. We need additional number patterns to compute these elements.

For the first step of the DMRG sweep, additional patterns are added. These patterns satisfy
\begin{align}
N+1\le n_S \le \min(4,2N) \\
0\le n_D \le 4 \\
n_S +n_D +n_E =2N
\end{align}

For the last step of the DMRG sweep, additional patterns are also added. These patterns satisfy
\begin{align}
N\le n_E \le \min(4,2N) \\
0\le n_D \le 4 \\
n_S +n_D +n_E =2N
\end{align}

Because $S$ (or $E$) is composed of a single site in the first or last step of the sweep, the
expectation value of an operator with more than two $a$'s or two $a^\dagger$'s is zero. Finally, the number of indices on single sites should not exceed 4.

\section*{Appendix B: Restriction on type patterns}
It is a waste of resources to compute RDM elements which are zero or redundant. Further RDM elements can be computed through
several choices of combinations of operators. We implemented restrictions to skip the calculations of zero and redundant RDM elements.

The operators $a_i$ and $a^\dagger_j$ are first permuted, so that the orbital indices do not decrease. Only permutations between two operators with different orbital indices are needed.
If $a_i$ precedes $a^\dagger_j$ and $i=j$, the corresponding compound operators are not required in the RDM calcualtion.
For a single site block, the compound operator with $a$ ahead of $a^\dagger$ is omitted. For other blocks, the loop over $o_io_jo_k\dots$ should satisfy $i \le j \le k\dots$ ('$=$' is only valid when the two adjacent operators are in normal order).

Further, the RDM (but not the transition RDM) is a Hermitian matrix. Using the 3-RDM as an example
\begin{equation}
  \langle a_i^\dagger a_j^\dagger a^\dagger_k a_l a_m a_n\rangle = \langle(a_i^\dagger a_j^\dagger a^\dagger_k a_l a_m a_n)^\dagger\rangle
  = \langle a_n^\dagger a_m^\dagger a_l^\dagger a_k a_j a_i\rangle
\end{equation}
We can always use transposes and permutations to ensure that the operator string begins with $a^\dagger$ rather than $a$.
Therefore, all type patterns begining with $a$ are omitted.

Similarly, if an operator begins with $a^\dagger_ia_i$, $a^\dagger_ia_ia^\dagger_ja_j$ or $a^\dagger_ia_ia^\dagger_ja_ja^\dagger_ka_k$, the remaining set of operators also has transpose and permutation symmetry. e.g.:
\begin{equation}
\begin{aligned}
  &\langle a^\dagger_ia_i(o_j o_k\dots)\rangle \\
  = &\langle(a^\dagger_ia_i (o_j o_k\dots))^\dagger\rangle \\
  = &\langle(o_j o_k\dots)^\dagger(a^\dagger_ia_i)^\dagger\rangle \\
  = &\langle a^\dagger_ia_i(o_j o_k\dots)^\dagger\rangle
\end{aligned}
\end{equation}
where $i < j\le k\dots$.
Thus, there is no need to compute type patterns where $a$ follows $a^\dagger_ia_i$.


\begin{acknowledgement}
This work was supported by the US Department of Energy. Primary
support was provided by DE-SC0010530 and additional support
was provided by DE-SC0008624.
\end{acknowledgement}

\providecommand{\latin}[1]{#1}
\providecommand*\mcitethebibliography{\thebibliography}
\csname @ifundefined\endcsname{endmcitethebibliography}
  {\let\endmcitethebibliography\endthebibliography}{}

\end{document}